\documentclass{article}

\usepackage{arxiv}

\usepackage[utf8]{inputenc} % allow utf-8 input
\usepackage[T1]{fontenc}    % use 8-bit T1 fonts
\usepackage{hyperref}       % hyperlinks
\usepackage{url}            % simple URL typesetting
\usepackage{booktabs}       % professional-quality tables
\usepackage{amsfonts}       % blackboard math symbols
\usepackage{nicefrac}       % compact symbols for 1/2, etc.
\usepackage{microtype}      % microtypography
\usepackage{lipsum}		% Can be removed after putting your text content
\usepackage{graphicx}
\usepackage{natbib}
\usepackage{doi}

\title{\emph{Hyperwave}: Hyper-Fast Communication within General Relativity}

\author{{\hspace{1mm}Lorenzo Pieri}
\\
	Createc LTD, UK \\
	Oxford Brookes University, UK\\
	\texttt{lorenzo.pieri@createc.co.uk, lpieri@brookes.ac.uk} \\
}

\renewcommand{\headeright}{ }
\renewcommand{\undertitle}{ }
\begin{document}
\maketitle

\begin{abstract}
Warp-drives are solutions of general relativity widely considered unphysical due to high negative energy requirements. While the majority of the literature has focused on macroscopic solutions towards the goal of interstellar travel, in this work we explore what happens in the small radius limit. In this regime the magnitude of the total negative energy requirements gets smaller than the energy contained in a lightning bolt, more than 70 orders of magnitude less than the original Alcubierre warp drive. Such an amount could conceivably be generated with current technology by scaling up Casimir-like apparatuses. We then describe a tubular distribution of externally-generated negative energy which addresses the major issues plaguing macroscopic warp-drives and propose a concrete mechanism to accelerate and decelerate a warp. A byproduct of warp deceleration is the emission of a ray of high-energy particles. The detection of such particles could be used as the backbone of a faster-than-light communication device, reminiscent of the “Hyperwave” of science fiction, even though significant engineering challenges remain to achieve practical communication.     
\end{abstract}

% keywords can be removed

\section{Introduction}
The seminal paper \citep{Alcubierre_1994} introduced in general relativity the idea of travelling faster than light while locally remaining on a timelike trajectory. These solutions, called warp drives due to the obvious application to space travel, are currently considered unphysical due to a series of shortcomings including the requirement of large quantities of negative energy densities. Further problems include requiring tachyonic matter and causal disconnection between interior and exterior for superluminal bubbles. 

Having the application of space travel in mind, the large majority of warp drives literature has focused on macroscopic solutions. Recent attempts aim to retain warp-like features with positive energy solutions \citep{lentz2021hyperfast}, but are still debated \citep{Santiago_2022}. 
In this paper we take a different approach: we embrace the negative energy requirements of warp geometries, but aim to minimise them by exploring applications in the small radius limit. We also approach the problem with a different mindset: can we take advantage of warp drive geometries without regard for the initial payload, which may be hard to engineer? It turns out that the answer is yes, thanks to the peculiar interaction properties of warps with matter: we can use warp drives as superluminal information carriers. We call these Hyperwaves.

We also show that the core problems affecting macroscopic warps are significantly mitigated for small bubbles suitable for communication. In particular we provide a concrete mechanism to physically accelerate warp drives, involving a tube-like region of negative energy which is engineered according to a velocity profile set in advance.

Even though for simplicity we select a warp metric which is known to be non-optimal in terms of negative energy requirements, we show that the total energy requirement for Planck-sized small bubbles can be satisfied with a Casimir device of reasonably large size. This paper therefore contributes to make warp drive physics one step closer to being an experimental science, even though we briefly mention in the conclusions that significant unsolved engineering challenges remain, together with considerations around causality and closed timelike curves.

The outline of the paper is the following: In the first chapter we recap the previous shortcomings found in macroscopic warp drive. In the second chapter we show how these problems disappear for a specific configuration usable for superluminal communication, that we call Hyperwave. We conclude by mentioning some remaining problems and promising areas of investigation.

\section{Previous literature results: Warps Basics and Shortcomings}

Warp drives geometries are solutions to the equation of General Relativity with a peculiar property: a bubble-like space-time perturbation propagates at arbitrary speed, while observers inside the bubble never locally move faster than light. In contrast to wormhole geometries, no topology changes are needed. In this paper we will use the words “warp bubble”, “warp drive”, “warp”, “bubble”, “drive” as synonyms. The “spaceship” is the observer at the centre of the warp bubble, which purportedly controls the drive (we will actually see that this cannot be the case).

The original metric introduced by \citet{Alcubierre_1994} is given by 

\begin{equation}
ds^2 = -dt^2 + (dx - v_s f(r_s) dt)^2 + dy^2 + dz^2 
\end{equation}

where we assume that our spaceship moves along the $x$ axis of a Cartesian coordinate system,   $x_s(t)$ is the spaceship trajectory described by an arbitrary function of time, and 

\begin{equation}
v_s(t) = \frac{dx_s(t)}{dt}
\end{equation}

\begin{equation}
r_s(t) = [(x - x_s(t))^2 + y^2 + z^2]^{1/2}
\end{equation}

and $f(r_s)$ is a function similar to a top hat, going to 1 for $r_s < R$, where $R$ is the bubble radius, and going rapidly to zero for $r_s > R$.

 For the spaceship trajectory we have  

\begin{equation}
	d\tau = dt
\end{equation}

that is proper time is equal to coordinate time inside the bubble, and since the coordinate time is the same as the proper time of a distant Minkowskian observer, the spaceship suffers no time dilation effect. Considering two stars A and B with proper distance D and an observer located in A, we are interested in measuring how much time it takes for the drive to go from A to B and back. In this way the warp drive will have travelled a proper distance D (observer independent quantity) in a proper time $\tau$
 (again, observer independent) as measured by the local observer. In special relativity we know that the time measured by the person making the round trip can be made arbitrarily small if his speed approaches $c$. However, the fact that in general relativity one can actually make such a round trip in an arbitrarily short time as measured by an observer that remained at rest is the non trivial feature of the warp drive. Then it can be shown \citep{Alcubierre_1994} that

\begin{equation}
t \simeq \tau \simeq 2 \sqrt{\frac{D}{a}}
\end{equation}

which is a coordinate independent statement showing that the measured time can be arbitrarily short, since $a$ (the warp acceleration) is arbitrary. A pedagogical reference covering warp drive basics is \citep{2017}.

Unfortunately for aspiring space travellers, it was immediately noticed that warps violate all the classical energy conditions and that the energy density seen by Eulerian observers is completely negative. Following the notations of \citep{Pfenning_1997}, the total energy needed to sustain a drive is given by:

\begin{equation}
E = - \frac{1}{12} v^2  (\frac{R^2}{\Delta} + \frac{\Delta}{12} ) 
\label{e}
\end{equation}

where $\Delta$ is the bubble wall thickness. By itself this is bad news but not a deal breaker, since we know that negative energy is common in quantum systems, resulting in physical systems breaking WEC, NEC and SEC \citep{Curiel_2017, Kontou_2020}. 

But in the context of Quantum Field Theory in curved space, Quantum Inequalities severely constrain the amount of negative energy that can be seen for a given amount of time \citep{Pfenning_1997,fewster2012lectures}
Applied to warps, they require the bubble walls to be extremely thin, close to Planck scale. Crunching the numbers, the total amount of negative energy required is north of $10^{80}$ Joule for a warp of radius $100 m$ moving at $1.1 c$, greater than the estimated total mass–energy of the observable universe.  

It should be emphasised that Quantum Inequalities have not been proven for realistic interacting quantum fields, where the they are expected to be weaker if they exist \citep{cadamuro2019energy}, and they often require Minkowskian conditions at infinity, a condition which is not appropriate for fields inside bounded regions, as in a Casimir devices. Nevertheless in this work we will assume that they do extend to more realistic scenarios and that they do limit the magnitude and duration of the negative energy density seen by an inertial observer to an amount similar to the QI used in \citep{Pfenning_1997}\footnote{This seems the right place to encourage further research on the validity of QIs for realistic scenarios, as this could have direct dramatic consequences on warps energy considerations.}. 

Following work by \citet{Broeck_1999} reduced dramatically the energy requirements, even though they still remained large in absolute sense. More papers looking into energy conditions and the sign of the energy density include \citep{Olum_1998, gauthier2002new, Santiago_2022}

It has been noticed that superluminal warp drives form black hole horizons, effectively causally disconnecting the interior with the exterior of the bubble \citep{Finazzi_2009}. In particular, it is not possible for the spaceship to steer the bubble.

In general relativity the stress energy tensor shapes the space-time, which then influences matter, and so on. It is often said that the warp drive is a case of bad metric engineering, in which the metric itself was designed with a purpose, but the derived energy tensor is unphysical. Indeed it has been noted that to move faster than light the warp geometry the negative energy sourcing the drive needs to be able to keep up with the drive \citep{broeck2000impossibility}. So superluminal matter is needed in the first place to source a warp drive.

In semiclassical gravity it is hinted that spaceships would experience an uncomfortable intense thermal flux at Planck temperature  \citep{Hiscock_1997, Finazzi_2009}. Moreover, the bubble itself seems to be unstable due to divergences in the renormalized stress-energy tensor at the horizons. Luckily recent works show that the problem seems not too severe for realistic spacetime dimensions \citep{Barcel__2022}, so we will not consider this issue here.

Finally, a perhaps unappreciated issue with warps is that no physical mechanism for acceleration has been provided. The warp drive in the original solution magically postulates a velocity function with associated arbitrary acceleration, but since no traditional “internal” propulsion applies, how is the drive achieving such acceleration? Related to this \citep{Bobrick_2021}, since the metric is time-dependent, how is the drive sourcing the additional energy required to maintain the drive? Indeed since the energy density needed to sustain the warp is related to the warp velocity, accelerating means varying the amount of energy density sourcing the drive. But this is impossible in traditional drives for the aforementioned reasons (the ship inside the bubble cannot control the bubble, to change the energy density superluminally one would need tachyons). In this sense a valid mechanism to provide acceleration is missing in literature, which is usually assuming an isolated superluminal bubble.

\section{Hyperwave}

In this chapter we will explain how to overcome the mentioned limitations of warp drives. As we will see, some trade-offs will need to be made, resulting in very small warps confined to tubular regions, moving with a predetermined path and velocity profile. While these downsides preclude using these devices for practical interstellar travel, they open up the potential for faster-than-light communications. Science fiction writers such as Asimov (“Foundation series”, 1942) used the name Hyperwave to denote faster-than-light communications. As such, we will refer to the fast moving warp bubbles as Hyperwaves, and (creatively) to the energy density tubes used to guide the Hyperwaves as Hypertubes.

\subsection{The Small Radius Limit}

The total negative energy requirement grows like the square of the bubble radius, so how small can we get if we are not interested in engineering the warp payload in advance? It turns out that to minimise E both terms of (\ref{e}) are required, since the radius and the bubble thickness are of similar size for Planckian bubble radii. In SI units:

\begin{equation}
E \approx - 10^{43} \cdot v_c^2  (\frac{R^2}{\Delta} + \frac{\Delta}{12} ) 
\end{equation}

where $v_c$ is the velocity in units of $c$.
For a bubble travelling at $v = 2 / \sqrt{3} c\approx 1.15c$, radius $R = 1.01L_p$, where $L_p \approx 1.6 \cdot 10^{-35} m$ is the Planck length, and $\Delta =  L_p$ we obtain that $E$ can be as low as

\begin{equation}
E \approx -2 \cdot 10^{8} J
\label{min_wave}
\end{equation}

less (in absolute magnitude) than the energy contained in an average lighting bolt or in about $10 kg$ of crude oil. One can check that $\Delta$ is well inside the requirements from the Quantum Inequalities and that all the solution features are equal or above the Planck scale. Indeed, following \citep{Pfenning_1997}, the minimum scale is $r_{min} \approx \frac{2 \Delta}{ \sqrt{3}  v_c} \approx  L_p$. As in Alcubierre, all point-like energy conditions are violated, but the violations will be localised given the small size of the bubble radius and bubble thickness.

In practice it may not be convenient to aim for the smallest possible radius, since as we will see the engineering required to stabilise and detect warps gets harder as the radius gets smaller.

How do we obtain negative energy in the first place? Negative energies are commonly found in many quantum systems, and famously in Casimir devices and squeezed quantum states. The Casimir effect, which has been experimentally confirmed multiple times \citep{PhysRevLett.78.5, Bressi_2002, Wang_2021}, borrows negative energy from the vacuum thanks to the boundary conditions to the quantum fields imposed by two plates, usually metallic. The magnitude of the Casimir effect is tiny:

\begin{equation}
E_{cas} = - k \frac{A}{d^3} 
\end{equation}

where

\begin{equation}
k = \frac{\hbar c \pi^2}{720}
\end{equation}

for instance for two metallic plates of area $A = 1 m^2$ at a distance of 1 micron we find $E_{cas} \approx -4 \cdot 10^{-10} J$.

With traditional materials it should be possible to push Casimir plates to operate at distances of few nanometers, but not much below since metallic plates are not reflective to very high frequency radiation, so the minimum distance is close to plasma wavelength of the metal or the transition wavelength of the atoms \citep{doi:10.1142/9383}. If we consider a 5-nanometer separated Casimir plates we get $E_{cas} \approx -3 \cdot 10^{-3} J$. Therefore we see that the required combined area of the plates necessary to obtain enough negative energy to produce an Hyperwave of energy (\ref{min_wave}) is of the order of a square plate of $100 km$ side, which is a large amount, but non inconceivably large given the potential technological impact. 

Notice that the stress-energy tensor induced by a Casimir device is diagonal, but the Alcubierre stress-energy tensor has in general non trivial momentum-density and momentum-flux components too. Purely considering a Casimir-like device, a stress-energy tensor momentum-density can be obtained by a Lorentz boost of the Casimir plates (or boosting the electric and magnetic fields around the plates) \citep{gorban2023lorentz}. A momentum-flux can be obtained by considering Casimir plates inside a non trivial magnetodielectrics medium 
 \citep{Philbin_2011}. These components can be used to engineer the bubble's direction of travel, but a proper treatment of the matter is beyond the scope of this work. Finally one can show that even in the presence of dielectrics there are regions of negative energy density \citep{Sopova_2002}.

To be clear, we are not implying that Casimir-like devices are the best energy source candidates since, as discussed in the conclusions, they provide low energy density. A further potential issue for Casimir-devices is that, contrary to superluminal solutions, Casimir-devices are expected to respect the achronal ANEC \citep{Graham_2007,Kontou_2015}, but those expectations are based on free-field calculations and it is not clear if they will keep doing so when more realistic interacting quantum fields are considered. Yet, the Casimir example is relevant, since it falsifies a commonly held belief in the warp literature, which is about warp geometries having an unphysically-high total negative energy requirement for any practical use-case. This is often used as the number one argument to dismiss the feasibility and general technological usefulness of warp geometries. 

It must be stressed that the nature of gravitational effects of negative energy from quantum systems is still an area of active investigation \citep{helfer1998physics, Fulling_2007}, featuring perhaps the worst theoretical prediction in the history of physics, the Cosmological Constant Problem \citep{Martin_2012}. Moreover there is not even a widely accepted definition for gravitational energy in the classical setting, where negative energies appear \citep{ha2005gravitational}.

The large negative energy requirements addressed here is perhaps the flashiest argument to dismiss macroscopic warp drives quoted in literature, but in reality two much more fundamental issues are present: how do we accelerate a warp and how do we avoid the catch-22 of superluminal matter to obtain superluminal motion? We cover this next.

\subsection{Acceleration}

Traditional objects are propelled in space-time by the ejection of propellant, exploiting the conservation of the total momentum. But warps are a different beast, being a non-trivial change in space-time itself. To make a warp accelerate, we will instead architect the energy distribution, the right-hand side of the Einstein equations.  

As we said, warps do form a black and white horizon and as such cannot be controlled from the inside. But nothing prevents us from controlling the warp from the outside, dynamically modifying the negative energy distribution to make the warp move. An additional benefit of controlling the warps from outside is that we can control and contain potential warp instabilities. We can imagine surrounding the drive by a Casimir device to alter the energy distribution while the drive is passing by, but for interesting drives we run into the tachyonic problem: how to make the energy distribution keep up with a superluminal drive, if the energy distribution is not superluminal in the first place? Getting superluminal rides using superluminal matter doesn’t sound that impressive… 

We can overcome this issue by building the energy distribution in advance! If we imagine being able to locally control the energy density in a tubular region of space, we can selectively activate the energy density of a rolling region of the tube to allow the drive to pass by, ensuring that the Hyperwave is inside this powered tube section at all times. By turning on and off the energy density in advance on the desired sections of the tube expected to host the Hyperwave at a given time, the hyperwave can be sourced even if superluminal. In each local region, we can concentrate the energy density into a spherical region\footnote{More precisely, for Alcubierre solutions the energy density is spread into a toroidal region perpendicular to the direction of motion.} of radius R to constraint the energy in the warp geometry to a given velocity v. For a warp moving only in the x axis, we can increase the magnitude of the energy density as x increases without modifying the spatial distribution of energy density, to increase the velocity of the warp while keeping the bubble radius fixed. So in any region of space we only need to support negative energy densities for a very short time, indeed the lower the better for our gravitational-energy bill \footnote{One may try to argue with the utility company that the bill should report a credit, since we are asking for negative energy.}. If the energy density varies slowly along the x axis, we can use (\ref{e}) to estimate the total energy needed to sustain a drive at constant v in a small region. Therefore the average acceleration between the two points 1 and 2 is given by 

\begin{equation}
a_{av} = [\frac{1}{12} (\frac{R^2}{\Delta} + \frac{\Delta}{12} )]^{-1/2} \frac{\sqrt{|E_2|} - \sqrt{|E_1}|}{T}
\end{equation}

where $T$ is the time taken from the warp to go from 1 to 2. Notice that for all the subluminal tracks, such as the initial acceleration and the final deceleration, there is no need to fix in advance the energy distribution. 

Architecting the energy density externally also solves the problems raised in \citep{Bobrick_2021} about energy conservation and the provenance of the energy needed to sustain the drive. Finally, it's worth commenting that different observers may see different signs of the energy source in Alcubierre, as shown in \cite{Carneiro_2022} for the static external observer, the latter being perhaps the most natural observer in a spacetime with a microscopic Hyperwave. 

Related to the Hypertube setup, engineering in advance the energy distribution for one-way superluminal trips has been already suggested in \citep{Krasnikov_1998, Everett_1997}, but deemed uninteresting for the sake of space travel. As an alternative, Krasnikov proposed a two-way design, the Krasnikov-tube, allowing arbitrary superluminal round trips.     

To recap, we have a highway of negative energy built in advance and featuring an arbitrary fixed-in-advance acceleration\footnote{Acceleration is not strictly needed. That said, physically realisable Hyperwaves would likely require an acceleration phase to go from subluminal to superluminal and a final deceleration phase.} and velocity profile. Having a fixed velocity profile is not an issue for a communication device, if anything this is a desired feature to make the system standardised. But can we perform arbitrary transmission of data with this device? By arbitrary transmission we mean that even though the velocity and energy profile is fixed, the information content of the Hyperwave is not.
Contrary to the macroscopic case, a peculiarity of the very small regime is that no conscious being, computer or storage device can fit inside such a narrow space. Moreover, a bubble may experience strong perturbations internally, such as strong thermal flux of particles, polluting our message. So how do we practically encode information using a warp, without dealing with the hard task of creating a microbubble which encapsulates that information? 
Some interesting options are to encode a bit of information into the shape of the warp or introducing a non trivial angular momentum. Unfortunately these options would need to alter the default energy distributions, but since we have fixed the energy distribution in advance they are ruled out. 

As we will see next, exploiting the interaction of the warps with other matter there is a way to encode bits of information into warps without significant modifications to the energy distribution.

\subsection{Detection}

Warps interaction with matter is a scarcely studied topic, yet some works that deal with it include \citep{Pfenning_1997, Schuster_2023}. The work \citep{McMonigal_2012} performs some numerical analysis to observe the propagation of null particles and matter when interacting with a warp.

An interesting property is the behaviour of particles which move in the same direction of the drive and get later captured by it. When particles are engulfed by the fast bubbles they find themselves locally in flat space and instead of going towards the centre of the bubble they keep moving towards the same region from which they entered. But as explained, for superluminal bubbles such particles encounter the bubble horizon and are unable to escape. Now the interesting part: if the bubble decelerates enough the horizon disappears, and all the collected particles are violently ejected in the original direction of motion. Moreover the particle's energy grows proportionally to the time spent in the bubble, resulting in a strong blueshift as seen by an external stationary observer \cite{McMonigal_2012}. Here we assume that the test particles do not significantly modify the warp energy density, which is justified for instance for a ray of photons and given the high energy density involved in the warp walls. 

While intense rays of particles being emitted towards the destination can be seen as a further reason to be sceptical about the practicality of macro warps, in the micro world this becomes a feature, since we can use the ejected particle as an experimental signature of the warp passage. 

Consider a future Hyperwave sender, Alice, interested in transmitting a single bit of information. The case for full-blown communications would trivially follow by repeating this procedure multiple times. Alice would activate a switch to release a dense beam of particles in the direction of the Hyperwave velocity, resulting in the Hyperwave scooping up some of the particles, bringing them onboard for the trip and releasing them with higher energy when the bubble decelerates. 
The Hyperwave receiver, Bob, will measure this excess of particle emission and annotate that a “1” bit has been transmitted. A warp arrival, at the time expected by the fixed velocity profile, without the associated particle's excess detection is instead encoded in a “0” bit.  

Alice and Bob trading a single warp for a single bit (or perhaps even more than one warp for transmission redundancy) is far from being an efficient communication protocol, but it proves the point that useful communication can be achieved. 

Given Hypertubes of equal length, Hypertubes with smaller radii require denser media to obtain the same particle throughput. More generally, different path shapes can be constructed to obtain the desired throughput rate over a given distance and velocity profile. It may be convenient to encode different particle detection levels to a larger alphabet than binary, including symbols such as “communication start/end” or “Null”, the latter indicating that no information is being transmitted. Having the Hypertube available for immediate use means having “Null” Hyperwaves being sent at all times according to the fixed velocity profile when communication is not needed. When communication is desired Alice can turn on the particle ray and encode a bit into a Hyperwave.
Of course nothing stops us from shutting off the Hypertube when not needed for a long period of time, but to bring it on again a light signal must travel the whole Hypertube to set up the Hyperwave's path in advance. We leave to future works the design of practical Hyperwave communication protocols.

\section{Conclusions}

In this work we revisited the major shortcomings of warp drive geometries and noticed that these can be mitigated in the small radius regime, given a proper configuration of the negative energy distribution. Such tubular distribution, dubbed Hypertubes, provide a concrete mechanism to accelerate and decelerate warp bubbles, which in this context are called Hyperwaves. 	

Superluminal communication or even fast subluminal communication are natural applications for the Hyperwave. The feasibility of long distance communication, perhaps even at interplanetary or interstellar distance, will depend on the feasibility of keeping energy requirements low by constructing devices able to generate short burst of energy. On a smaller scale, it is tantalising to consider the fabrication of microchips capable of superluminal computing.

Even though this work outlines some positive results towards establishing the physics of warp drive spacetimes, some major “engineering” issues do remain. Firstly, we discussed how Hyperwaves total negative energy requirements can be up to 70 orders of magnitude less stringent than the original Alcubierre drives, but we did not discuss the magnitude of the energy density itself. Being the energy focused in a tiny region, the negative energy density involved is enormous, so even though we may be able to produce enough energy in future mega projects we must figure out how to focus them. In this regard, Quantum Energy Teleportation \citep{hotta2011quantum, Funai_2017}, which has recently been experimentally demonstrated \citep{Rodr_guez_Briones_2023}, could be of help. Secondly, closely related to the first issue, extremely fine manipulations of negative energy densities will be required to architect, operate and maintain Hypertubes. Finally, a more theoretical issue: as in Alcubierre warp drives \citep{PhysRevD.53.7365}, wormholes and Krasnikov tubes \citep{Everett_1997}, we  suspect that standard tachyonic-antitelephone-like arguments can be used to engineer a web of Hypertubes to build closed timelike curves. Whether this should be considered of  relevance is left to the reader's discretion; we limit to notice that no decisive argument can be made against CTCs without a quantum gravity theory \citep{visser2002quantum}.     

Speaking of quantum gravity, it must be stressed that the following work was carried inside the framework of general relativity, but for very small Hyperwaves we can expect quantum gravity corrections to be relevant. 

Regarding further areas of exploration, in this work we used the Alcubierre metric to illustrate the concept of Hyperwaves for simplicity and due its popularity, but this is perhaps the worst possible metric in terms of energy requirements, requiring the stress energy tensor to be negative everywhere. More optimised metrics could be considered instead, allowing to consider larger bubble radii, therefore minimising engineering challenges for a first experimental setup. Similarly, a Casimir device with large overall surface area has been considered as a source of negative energy, but more practical and energy-dense sources should be researched. Longer term, designing practical Hyperwave communication protocols will be needed. More generally, further understanding of the gravitational effects of negative  energy from quantum systems is an important theoretical area of investigation. 	

\section{Acknowledgements}

This work was partially supported by the Defence Science and Technology Laboratory (DSTL) and by the United Kingdom Ministry of Defense (MOD). 

Thanks to K. Olum, C. E. Gauthier, A. Bobrick, E. Lentz for useful conversations or shared references. 

I am grateful to the anonymous referees for important insights, suggestions and recommendations that improved the paper.

\bibliographystyle{unsrtnat}
\bibliography{references}  %%% Uncomment this line and comment out the ``thebibliography'' section below to use the external .bib file (using bibtex) .





% \end{thebibliography}

\end{document}